\documentclass[aps,pra,preprint,amsmath,amssymb]{revtex4}

\usepackage{graphicx,epsfig,epsf}
\usepackage{amsfonts}

\begin{document}
\newcommand{\ri}{{\rm i}}
\newcommand{\re}{{\rm e}}
\newcommand{\bx}{{\bf x}}
\newcommand{\br}{{\bf r}}
\newcommand{\bk}{{\bf k}}
\newcommand{\bE}{{\bf E}}
\newcommand{\bR}{{\bf R}}
\newcommand{\bn}{{\bf n}}
\newcommand{\rSi}{{\rm Si}}
\newcommand{\beps}{\mbox{\boldmath{$\epsilon$}}}
\newcommand{\rg}{{\rm g}}
\newcommand{\tr}{{\rm tr}}
\newcommand{\xmax}{x_{\rm max}}
\newcommand{\ra}{{\rm a}}
\newcommand{\rx}{{\rm x}}
\newcommand{\rs}{{\rm s}}
\newcommand{\rP}{{\rm P}}
\newcommand{\up}{\uparrow}
\newcommand{\down}{\downarrow}
\newcommand{\hc}{H_{\rm cond}}
\newcommand{\kb}{k_{\rm B}}
\newcommand{\cI}{{\cal I}}
\newcommand{\cE}{{\cal E}}
\newcommand{\cC}{{\cal C}}
\newcommand{\oF}{\overline{F}}
\newcommand{\Ubs}{U_{\rm BS}}
\newcommand{\qq}{{\bf ???}}

\sloppy

\title{Quantum Cloning without Interference} 
\author{Benoit Roubert and Daniel Braun}
\affiliation{Laboratoire de Physique Th\'eorique --- IRSAMC, UPS and CNRS,
  Universit\'e de Toulouse, 
  F-31062 Toulouse, FRANCE}

\begin{abstract}
We study the role of interference in the process of quantum cloning. We show
that in order to achieve better than classical cloning of a qubit no
interference is needed. In particular, a large class of symmetric universal
1$\to$ 2 qubit cloners exists which achieve the optimal average fidelity
$5/6$ for such machines, without using any interference. We also 
obtain 
optimal average fidelities for interference--free cloning in asymmetric
situations, 
and discuss the relation of the quantum cloners found to the
Bu\v{z}ek--Hillery quantum cloner. 
\end{abstract}

\maketitle

\section{Introduction} 
The celebrated no-cloning theorem \cite{Wootters82} states that unknown
quantum states cannot be perfectly copied, i.e.~no quantum
mechanical evolution exists  
which would transform a quantum state $\psi$ according to $\psi\rightarrow
\psi\otimes \psi$ for any unknown $\psi$. This theorem is at the basis of
the security of quantum key distribution systems, and has thus
tremendous technological implications. From the theoretical point of view,
it is a simple consequence of the linearity of quantum mechanical time
evolution. A corresponding theorem for classical probability
distributions can be proved just as easily: Propagating classical
probability distributions linearly (e.g.~phase space densities according to the
Liouville equation) does not allow to transform any unknown probability
distribution $p$ according to $p\rightarrow p\otimes p$. So how could you
download a perfect copy of this article? 

Classical and quantum information differ fundamentally by the fact that for
classical systems we have (at least in principle) unrestricted access to the
distributed quantities themselves (e.g.~positions and momenta of particles
in phase space), whereas for quantum systems we do not. In every--day
language, we understand under classical
copying measuring those quantities, and then
preparing another system with the same values of those quantities. Indeed,
a photocopy machine measures the position of bits of ink on the page,
assumes that their momenta are zero (but could in principle also measure
them), and
then prepares an arbitrarily large number of other pages with the same
positions of bits of inks with 
zero momentum. If we work with an ensemble of initial documents, differing
slightly in the positions of bits of ink, such that it is worth talking about a
probability distribution $p$, the latter will obviously be copied perfectly by
the copy machine as well if it copies each member of the ensemble perfectly
--- but the transformation 
is manifestly non--linear.

According to quantum mechanics, all information 
about a quantum system is encoded in the wavefunction $\psi$, and $\psi$
never attributes a sharp value to both positions and momenta of the 
particles involved. More importantly, the apparent failure of ``local hidden
variable'' descriptions of quantum mechanical correlations confronted with
experiments testing Bell-type 
inequalities  
\cite{Aspect82,Weihs98,Aspect99,Rowe01}
suggests that in a quantum mechanical system physical 
observables do not even have an objective value till the observable is
measured. Quantum
cloning aims therefore directly at reproducing the wavefunction rather than,
say, the coordinates and momenta of individual particles. If we want to
compare classical and quantum copying on equal 
footing, we should consider directly linear transformations (stochastic
maps) of the initial
probability 
distributions in the classical case, and this is what we
are going to do below.

Shortly after the publication of the no--cloning theorem, Bu{\v z}ek and
Hillery (BH) 
showed that pretty good approximate cloning is possible
\cite{Buzek96}. They invented a 
1 $\rightarrow$ 2 quantum copying machine which takes a first qubit $A$ in an
unknown pure state $|\psi\rangle=\alpha|0\rangle+\beta|1\rangle$ and a
second qubit $B$ in a fixed known 
initial pure state $|0\rangle$ 
(an ``empty page'') as input  and produces two mixed states
$\rho_A'=\rho_B'=\frac{5}{6}|\psi\rangle\langle\psi|+\frac{1}{6}
|\psi^\perp\rangle\langle\psi^\perp|$, where
$|\psi^\perp\rangle=\alpha^*|1\rangle-\beta^*|0\rangle$ is the state 
orthogonal to $|\psi\rangle$. A single ancilla qubit is necessary to perform
the transformation. The BH machine is manifestly symmetric, i.e.~the
fidelities $F_j$ obtained as overlap of the final mixed states with the
initial pure state of $A$, $F_j=\langle\psi|\rho_j'|\psi\rangle$, $j=A,B$, are
the same. It is also universal, in the sense that it copies all initial
states $|\psi\rangle$ with the same fidelity $F_j=5/6$. Later it was shown
that $5/6$ is indeed the optimal fidelity for symmetric universal
$1\rightarrow 2$ cloners of a qubit  \cite{GisinM97,Bruss98}, and that this
is the largest fidelity compatible with the non--signaling constraint imposed
by special relativity \cite{Gisin98}. 

For non--universal cloners, the fidelities
$F_j(\theta,\phi)=\langle\psi(\theta,\phi)| 
\rho_j' | \psi(\theta,\phi)\rangle$ depend on the initial 
states $|\psi(\theta,\phi)\rangle$, parametrized by polar and azimuthal
angles $\theta$ and $\phi$ on the Bloch sphere. It is then convenient to
introduce   
average  fidelities $\overline{F}_A=\overline{F}_B$, 
\begin{equation} \label{fids}
\overline{F}_j = \frac{1}{4\pi}\int_{0}^{2\pi}d\phi \int_{0}^{\pi}d\theta\,
F_j(\theta,\phi)\sin\theta\,. 
\end{equation}
The best possible {\em
  classical} cloning of a qubit with symmetric average fidelities leads to
$\overline{F}_A=\overline{F}_B=2/3$ \cite{Horodecki99}. 
The cloning is classical in the sense introduced above,
  i.e.~the cloner
acts with a classical stochastic map on the vector of probabilities
corresponding to the initial state, while it  has no access to
the quantum coherences. We will provide a new and simple demonstration of
  this result below.
The optimal classical fidelities are very easily matched by quantum cloning. A
  universal symmetric quantum cloner with 
fidelities $\overline{F}_A=\overline{F}_B=2/3$ can be constructed trivially by measuring
  the first 
  qubit in 
an orthogonal basis chosen randomly and uniformly over the Bloch sphere, and
preparing the second qubit in the state found in the measurement
  \cite{Scarani05}. Another trivial symmetric universal cloning machine
  leaves the 
original qubit unperturbed, produces the new one in a random pure state, and
swaps the two with probability 1/2. It has
fidelities $\overline{F}_A=\overline{F}_B=3/4$.  

As always, when a quantum protocol offers a higher performance than the best
possible classical scenario, it is natural to ask which quantum resource is
at the origin of this difference. ``Entanglement'' and ``Interference'' are
two obvious candidates \cite{Bennett00}. In \cite{Scarani05} it was argued
that quantum cloners without coherent interaction between original qubit and
copy should 
have fidelities limited by the one obtainable from the second ``trivial''
cloning strategy described above. Bru\ss{} and Macchiavello studied the
entanglement at the output 
of a universal quantum cloner (QC) and found that bipartite entanglement
between two outputs 
vanishes for $N\to M$ quantum cloners ($N$   
identical input qubits, $M$ output qubits) in the
``classical'' limit  $M\to\infty$ ($N$ fixed) of infinitely many copies, and
in fact as 
soon as $M\ge N+2$.
\cite{BrussM03}. The 
fact that 
classical cloning discards the quantum coherences whereas quantum
cloning need not, leads naturally to the
hypothesis that interference should play an important role in quantum
cloning. Put the other way round, it seems likely that a
cloner 
which uses no interference at all should not be able to outperform classical
cloning. Surprisingly, we show in this article that the contrary is true:
A large continuous class of quantum cloners exists which use strictly
zero interference, but which outperform classical cloning, and in fact give
the maximal possible average fidelity $5/6$ in the symmetric case. 

Evidently, in order to study this question, the concept of interference
needs to be made precise. A quantitative measure of interference was
introduced in \cite{Braun06}, and was used to study interference in quantum
algorithms. This led to the hypothesis that an exponential amount of
interference is needed in any unitary quantum algorithm which offers an
exponential speed up over the corresponding classical algorithm. The
hypothesis was supported by further numerical evidence in a study of
disturbed versions of Grover's   
and Shor's algorithms \cite{BraunG08}. In
\cite{Arnaud07} we demonstrated that the interference in a randomly
chosen quantum algorithm is with high probability close to the maximum
possible value, whereas in  \cite{Lyakhov07} we showed that interference
plays at most a minor role
in the transfer of quantum information through spin chains. 

Below we briefly
review the 
interference measure. We will then formulate quantum cloners very generally
in terms of a 
dynamical matrix \cite{Zyczkowski04}, and rewrite the interference measure
using the dynamical matrix. The average fidelities are linear
functions in the matrix elements, and it turns out that vanishing
interference leads to linear constraints. Finding the optimal average fidelity
becomes an instance of semi-definite programming, which we solve both for
given $\overline{F}_A$ or $\overline{F}_B$, and for
$\overline{F}_A=\overline{F}_B$. The latter case leads to one specific
quantum cloner, very similar to the BH quantum cloner, and in particular with
the same optimal average fidelities. We show, however, that starting from
that machine a whole continuous class of other machines can be constructed
with the same optimal average fidelities and vanishing interference. 

A final remark is in order concerning the applicability of our results to
quantum 
broadcasting. Sometimes the term ``quantum cloning'' is restricted to
machines which leave the outputs in a product state. ``Quantum 
broadcasting'' 
does not make any such restriction \cite{Barnum96}. In the following we 
will 
not make this distinction, i.e. we do not restrict ourselves to machines with 
factorizing outputs such that these machines might as well be called quantum
broadcasters.

\section{Interference measure}
The main physical intuition behind the interference measure is that it
should measure the coherence of a propagation, as well as equipartition: A
classical stochastic map (i.e.~a process which is not coherent at all),
should not give rise to interference, whereas a unitary evolution
might. However, a pure permutation of basis states would normally not
be considered as creating interference, even if completely coherent. Clearly,
initial states need to be ``split'' and superposed. The more states
contribute with appreciable amplitude to a final state, the larger the
interference. This is the ``equipartition'' property. In \cite{Braun06} we
derived an interference measure by studying the change of final state
probabilities as function of phases in the initial states. This led to an
interference measure which, while probably not unique, does 
measure both coherence and equipartition. It maps any propagator
$P$ of a density matrix $\rho$, $\rho'=P\rho$, to a real number between 0
and $N-1$, where $N$ is the dimension of the Hilbert space on which $\rho$
and $\rho'$ act. Note that
$P$ is a super-operator, which, when specified in the computational basis
$\{|i\rangle\}$,
propagates $\rho$ with matrix elements $\rho_{ij}=\langle
i|\rho|j\rangle$ according to
\begin{equation} \label{rhoij}
\rho_{ij}'=\sum_{k,l}P_{ij,kl}\rho_{kl}\,.
\end{equation}
In terms of $P$, the interference $\cI$ is given by \cite{Braun06}
\begin{equation} \label{IM}
\cI(P)=\sum_{i,k,l}|P_{ii,kl}|^2-\sum_{i,k}|P_{ii,kk}|^2\,.
\end{equation}
Clearly, if $P$ only propagates probabilities (i.e.~reduces to a classical
stochastic map), $P_{ii,kl}\propto \delta_{kl}\forall k,l$, where
$\delta_{kl}$ is the Kronecker-delta, $\cI(P)$ vanishes. The squares in
(\ref{IM}) assure the measurement of equipartition. This becomes most obvious
for unitary propagation, $\rho'=U\rho U^\dagger$, in which case $\cI$
reduces to $\cI=\left(N-\sum_{i,k}|U_{ik}|^4\right)$, where the double sum
is nothing but the sum of inverse participation ratios of the column vectors
of $U$.

\section{Quantum cloning in terms of the dynamical matrix}
\subsection{Dynamical matrix} 
We formulate 1$\to$ 2 qubit quantum cloning very generally in terms of the dynamical
matrix $D$ \cite{Choi75,Zyczkowski04}. $D$ is related to $P$ by a simple
reshuffling of 
indices, $D_{ij \atop kl}= P_{ik,jl}$, and therefore contains all
information about the propagation of the two--qubit system (original qubit
and copy). In terms of $D$, eq.(\ref{rhoij}) reads
\begin{equation} \label{rhoijD}
\rho_{ij}'=\sum_{k,l}D_{ik \atop jl}\rho_{kl}\,.
\end{equation}
The advantage of this is that if one considers $ik$ and $jl$ 
as single indices $I$ and $J$ (taking values $0,\ldots,15$ for
$ik=00,01,\ldots 03,10,11,\ldots,33$, and similarly for $jl$),
$D_{ik \atop jl}\equiv D_{IJ}$, 
the matrix $D$ acquires certain useful properties, inherited from the
properties of $\rho'$. It can be shown that $D$ is hermitian
($D_{IJ}=D_{JI}^*$) and block positive ($\langle \psi|D|\psi\rangle\ge 0$
$\forall |\psi\rangle =|u\rangle\otimes|v \rangle$, where $|u\rangle$ and
$|v\rangle$ are arbitrary pure single qubit states).    
Block-positivity implies immediately that all diagonal matrix elements of $D$
must satisfy $0\le D_{ij \atop ij}\le 1$. 
Furthermore,
\begin{equation}\label{pT}
\sum_{i=0}^3 D_{ij \atop ij}=1 \, \forall\, j
\end{equation}
assures the correct normalization of $\rho'$ \cite{Zyczkowski04}.
Given the linear nature of the propagation of the density matrix,
eq.(\ref{rhoijD}), the average over all initial states for the fidelities
$\oF_A$ and $\oF_B$ can be performed once and for all, and leads to 
\begin{eqnarray}
\oF_A&=&\frac{1}{6}\Big(
2D_{00 \atop 00}+D_{00 \atop 22}+D_{02 \atop 02}+2D_{10 \atop 10}+D_{10 \atop 32}+D_{12 \atop 12}\nonumber\\
&&+D_{20 \atop 20}+D_{22 \atop 00}+2D_{22 \atop 22}+D_{30 \atop 30}+D_{32 \atop 10}+2D_{32 \atop 32}\Big)\,,\label{Fad}\\
\oF_B&=&\frac{1}{6}\Big(
2D_{00 \atop 00}+D_{00 \atop 12}+D_{02 \atop 02}+D_{10 \atop 10}+D_{12 \atop 00}+2D_{12 \atop 12}\nonumber\\
&&+2D_{20 \atop 20}+D_{20 \atop 32}+D_{22 \atop 22}+D_{30 \atop 30}+D_{32 \atop 20}+2D_{32 \atop 32}\label{Fbd}
\Big)\,.
\end{eqnarray}
Therefore, only 12 matrix elements of $D$ determine the average fidelity of
the $A$ and $B$ clones. Equations (\ref{Fad},\ref{Fbd}) can be rewritten in
the form  
\begin{equation} \label{FAA}
\oF_A={\cal A}\cdot D,\,\,\,\,\oF_B={\cal B}\cdot D\,
\end{equation}
with two hermitian matrices ${\cal A}$ and ${\cal B}$ easily extracted from
eqs.~(\ref{Fad},\ref{Fbd}).  The multiplication in eq.(\ref{FAA}) is
understood as the 
scalar product between two $16\times 16$ matrices, i.e.~ 
\begin{equation} \label{sp}
{\cal A}\cdot D=\sum_{I,K=0}^{15}{\cal A}_{IK}\cdot D_{IK}\,.
\end{equation}
As for the interference, eq.(\ref{IM}) is rewritten as 
\begin{equation} \label{IMD}
\cI(D)=\sum_{i,k,l}|D_{ik \atop il}|^2-\sum_{i,k}|D_{ik \atop ik}|^2\,. 
\end{equation}
In order to have zero interference, it is then clear that we must have 
\begin{equation} \label{zI}
D_{ik \atop il}=0 \,\, \forall k\ne l\,.
\end{equation}
In other words, when $D$ is written as a matrix (with indices $I$, $K$
introduced above), the off-diagonal matrix elements of the diagonal blocks
of $D$ must vanish.

\subsection{Classical propagation}
As discussed in the Introduction, we understand under classical cloning of a
probability distribution 
the propagation of the probabilities with a stochastic classical map,  
\begin{equation} \label{rhoijC}
\rho_{ii}'=\sum_k D_{ik \atop ik}\rho_{kk}\,,
\end{equation}
 i.e.~only the diagonal matrix elements $D_{II}$ contribute. As we
 have furthermore eq.(\ref{pT}), we
 obtain  
the average classical fidelities 
\begin{eqnarray}
\oF_A^{cl}&=&\frac{1}{6}\left(3+D_{00 \atop 00}+D_{10 \atop 10}-D_{02 \atop 02}-D_{12 \atop 12}\right)\\
\oF_B^{cl}&=&\frac{1}{6}\left(3+D_{00 \atop 00}+D_{20 \atop 20}-D_{02 \atop 02}-D_{22 \atop 22}\right)\,.
\end{eqnarray}
Since $0\le D_{00 \atop 00}+D_{10 \atop 10}\le 1$, and $0\le
D_{02 \atop 02},D_{12 \atop 12}\le 1$, (as well as $0\le
D_{00 \atop 00}+D_{20 \atop 20}\le 1$, and $0\le 
D_{02 \atop 02},D_{22 \atop 22}\le 1$ concerning $\oF_B^{cl}$) we
obtain immediate   
bounds for  $\oF_A^{cl}$ and $\oF_B^{cl}$, 
\begin{equation} \label{bFcl}
\frac{1}{3}\le \oF_A^{cl},\oF_B^{cl}\le \frac{2}{3}\,.
\end{equation}
The upper bound reproduces the one found in \cite{Horodecki99}. Thus, unknown
pure states of a single qubit cannot be cloned 
classically with average fidelity better than $2/3$. Note that the upper
bound is saturated by a classical identity map
($D_{ik \atop ik}=\delta_{ik}$), which copies perfectly the
probabilities,  but not the quantum coherences.

\section{Optimization}
The average fidelities $\oF_A$ and $\oF_B$ are linear functions and
therefore convex in the matrix elements $D_{ik \atop jl}$. There are
12 real independent matrix elements $D_{ij \atop ij}$ on the diagonal,
and six $4\times 4$ sub-blocks with non-vanishing complex matrix elements in
the upper block triangle,
such that $D$ contains 204 real independent variables. If
we are interested in a symmetric cloner with $\oF_A=\oF_B$, or if we
want to find the maximum or minimum value of $\oF_A$ for given $\oF_B$ (or 
vice 
versa), we add another
linear constraint and effectively reduce the number of variables by
1. The optimization is over all dynamical matrices satisfying the above
mentioned constraints (hermiticity, block positivity and partial traces
equal one, as well as  vanishing off-diagonal matrix elements in the
diagonal blocks in order to have vanishing interference). The large number
of independent variables in the optimization problem  calls for a
numerical solution. Fortunately, the problem falls into the class of convex
optimization problems, as both the function to be optimized and the allowed
domain for $D$ is convex: it is very 
easy to show that the set of matrices which are block-positive is convex,
and all other constraints are linear, i.e.~do not change the convexity of
the allowed domain. Furthermore, we note that if we impose the stronger
constraint of positivity, the problem reduces to an instance of
semi-definite programming (i.e.~optimization of a convex function over a
positive hermitian matrix, with eventual additional linear
constraints). Routines for solving semi-definite programming problems are
readily available. While one might worry that imposing positivity instead of
block-positivity leads to a lack of generality, we show that for a
symmetric cloner with zero interference the maximum allowed average
fidelity $5/6$ is reached in the space of positive matrices, such that
extending 
the space of matrices 
to those which are block-positive but not positive does not improve the
result. In the following we will call ${\cal D}$ the convex set of all
positive dynamical matrices giving rise to zero interference. 

We used the ``sedumi'' routine of the YALMIP package under Matlab to solve
the semi-definite programming problem, and 
found that for the symmetric case $\oF_A=\oF_B$ the dynamical matrix $D^{\rm
  opt}$ whose 
only non--vanishing matrix elements $D^{\rm opt}_{IJ}$ are the diagonal ${\rm
  diag} \left(\frac{2}{3},\frac{1}{4},0,\frac{1}{4},
\frac{1}{6},\frac{1}{4},\frac{1}{6},\frac{1}{4},\frac{1}{6},\frac{1}{4},
\frac{1}{6},\frac{1}{4},0,\frac{1}{4},\frac{2}{3},\frac{1}{4} \right)$, 
the elements
$D_{0\,6}^{\rm opt}=D_{6\,0}^{\rm opt}=D_{0\,10}^{\rm opt}=D_{10\,0}^{\rm
  opt}=D_{4\,14}^{\rm opt}=D_{14\,4}^{\rm opt}= D_{8\,14}^{\rm
  opt}=D_{14\,8}^{\rm opt} 
=\frac{1}{3}$,  
and $D_{4\,8}^{\rm opt}=D_{8\,4}^{\rm opt}=D_{6\,10}^{\rm
  opt}=D_{10\,6}^{\rm opt}=\frac{1}{6}$, leads to maximum 
average fidelities $\oF_A=\oF_B=5/6$. This is the
maximum average 
fidelity possible for a symmetric universal cloner
\cite{GisinM97,Bruss98,Gisin98}. In 
particular, the maximum average 
fidelities are larger than the allowed classical value $2/3$, and we have
therefore demonstrated that better than classical quantum cloning is
possible without using interference. Further below we will investigate this
machine in more detail, and compare it in particular to the BH QC.
It is easy to check that the found QC is universal, i.e.~all pure
initial states are copied with the same fidelity $F_A=F_B=5/6$. The minimal
possible average fidelity 
of $\oF_A$ for unrestricted $\oF_B$ (and vice versa) is
found to be $1/3$.

\begin{figure}
\epsfig{file=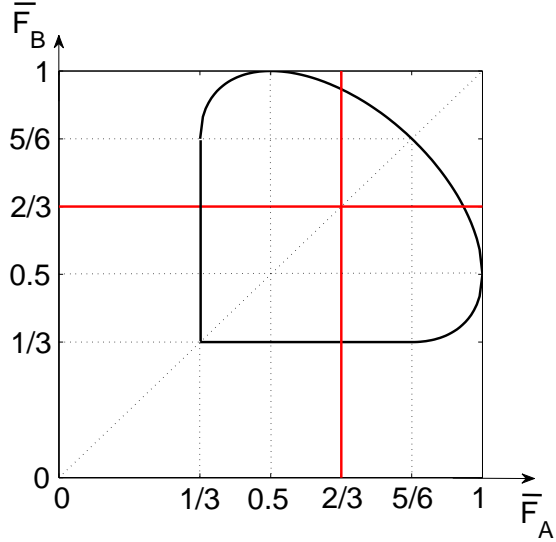,width=10cm,angle=0}
\caption{(Color online) The maximal and minimal average fidelity $\oF_B$ of
  an   interference 
  free  cloner as function of $\oF_A$ (black line) enclose the convex domain
  of all 
  possible interference--free cloners with positive dynamical matrix $D$. The
  maximum allowed classical fidelity for a symmetric cloner is
  $2/3$ (red lines).}\label{fig.1}         
\end{figure}

In the case of given $\oF_A$ (or $\oF_B$), we show in Fig.\ref{fig.1} the
maximum and minimum averaged fidelities $\oF_B$ (or $\oF_A$) as a function
of 
$\oF_A$ (or $\oF_B$) 
for the full allowed range of the latter. We see that $\oF_A$ can reach
unity, if at the same time the other qubit is completely randomized
($\oF_B=1/2$). The figure also shows that there is a whole continuous range
of asymmetric QCs where both $\oF_A$ and $\oF_B$ are larger than the
classical upper-bound$2/3$.  
The curves $\oF_A\left( \oF_B \right)$ and $\oF_B\left( \oF_A \right)$
enclose 
the domain af all possible 
QCs with positive dynamical matrix D. The domain is convex as $\oF_A$ and
$\oF_B$ are linear functions 
of the $D_{ik \atop jl}$ which form the convex domain
$\cal D$. 

\section{More general interference--free symmetric QCs}
Semidefinite programming guarantees that a local optimum is also a global
optimum. However, this does not exclude that other machines may
exist
with the same fidelity. Indeed, we shall show now that a high dimensional
continuous class of symmetric QCs exists which all reach maximum
average fidelity without making use of interference. 

First observe that the dynamical matrix $D^{\rm opt}$   has eigenvalues 1
(doubly degenerate), $1/4$ (8 times degenerate), and 0 (6 times
degenerate). Since $D^{\rm opt}$ is real and symmetric, its eigenvectors
form a real orthogonal matrix $O$. 
The idea is then to add a hermitian perturbation $W$ to $D^{\rm
  opt}$ in the 
subspace of eigenvectors of $D^{\rm opt}$ corresponding to the
non--vanishing eigenvalues, and which leaves the partial traces unchanged.
If the perturbation is small enough, the 
positivity of $D^{\rm opt}$ cannot be jeopardized immediately. Furthermore, a
perturbation can be easily constructed such that it is orthogonal to both
${\cal 
  A}$ and ${\cal B}$ in the sense of the scalar product (\ref{sp}). Let
$w_{IJ}$ be the matrix elements of $W$ ($0\le I,J\le 15)$ in the eigenbasis
of $D^{\rm opt}$. 
Transformed back to the computational basis, $\widetilde{W}=OWO^{T}$, we find 
${\cal A}\cdot \widetilde{W}$=${\cal B}\cdot \widetilde{W}
$=$5(w_{00}+w_{11})/12$.  
Thus we have unmodified success probability $\oF_A$ for 
\begin{equation} \label{oa}
w_{11}=-w_{00}\,.
\end{equation}
 From the requirement of unchanged partial traces we get
\begin{eqnarray}
w_{00}=w_{11}&=&0\\
w_{3\,3}+w_{5\,5}+w_{7\,7}+w_{9\,9}&=&0\\
w_{2\,2}+w_{4\,4}+w_{6\,6}+w_{8\,8}&=&0\label{w22}\,.
\end{eqnarray}
The
only thing which remains to be imposed is that transformed back to the
computational basis the perturbation still has vanishing off--diagonal matrix
elements on the diagonal blocks, in order to keep the interference at
zero. This requires $w_{0\,1}$,
$w_{0\,2}$,  $w_{0\,3}$, $w_{0\,4}$, $w_{0\,5}$, $w_{0\,6}$, $w_{0\,7}$,
$w_{1\,4}$, $w_{1\,5}$, $w_{1\,6}$, $w_{1\,7}$, $w_{1\,8}$, $w_{1\,9}$, 
$w_{2\,3}$, $w_{4\,5}$, $w_{6\,7}$, $w_{8\,9}$  to vanish. 
We are left with a large class of dynamical matrices $\widetilde{D}^{\rm
  opt}$ which can be written 
as a $4\times 4$ matrices of $4 \times 4$ complex sub-blocks
$\widetilde{D}^{\rm opt}_{I\,J}$ with $ \widetilde{D}^{\rm opt}_{I\,J} =
\left(\widetilde{D}^{\rm opt}_{J\,I}\right)^{\dag}$  and  
$ I,J \in \left\{ 0 ,\ldots , 3 \right\} $ where $\widetilde{D}^{\rm
    opt}_{0\,0} = {\rm diag}
\left(\frac{2}{3},\frac{1}{4}+w_{9\,9},0,
\frac{1}{4}+w_{8\,8}\right)$,  
$\widetilde{D}^{\rm opt}_{1\,1} =
  {\rm diag}\left(\frac{1}{6},\frac{1}{4}+w_{7\,7},\frac{1}{6},
\frac{1}{4}+w_{6\,6}\right)$, $\widetilde{D}^{\rm opt}_{2\,2} =
  {\rm diag}\left(\frac{1}{6},\frac{1}{4}+w_{5\,5},\frac{1}{6},
\frac{1}{4}+w_{4\,4}\right)$, $\widetilde{D}^{\rm opt}_{3\,3} =
  {\rm diag}\left(
0,\frac{1}{4}+w_{3\,3},\frac{2}{3},\frac{1}{4}+w_{2\,2} \right)$
and    
\begin{equation} \label{}
\widetilde{D}^{\rm opt}_{0\,1}=
\left(
\begin{array}{cccc}
0&0&\frac{1}{3}&0\\
\frac{w_{0\,9}^{*}}{\sqrt{6}}&w_{7\,9}^{*}&0&w_{6\,9}^{*}\\
0&0&0&0\\
\frac{w_{0\,8}^{*}}{\sqrt{6}}&w_{7\,8}^{*}&0&w_{6\,8}^{*}\\
\end{array}
\right), \,\,\,
\label{}
\widetilde{D}^{\rm opt}_{0\,2}=
\left(
\begin{array}{cccc}
0&0&\frac{1}{3}&0\\
\frac{w_{0\,9}^{*}}{\sqrt{6}}&w_{5\,9}^{*}&0&w_{4\,9}^{*}\\
0&0&0&0\\
\frac{w_{0\,8}^{*}}{\sqrt{6}}&w_{5\,8}^{*}&0&w_{4\,8}^{*}\\
\end{array}
\right)\,,
\end{equation}

\begin{equation} \label{}
\widetilde{D}^{\rm opt}_{0\,3}=
\left(
\begin{array}{cccc}
0&\sqrt{\frac{2}{3}}w_{1\,3}&0&\sqrt{\frac{2}{3}}w_{1\,2}\\
0&w_{3\,9}^{*}&\sqrt{\frac{2}{3}}w_{0\,9}^{*}&w_{2\,9}^{*}\\
0&0&0&0\\
0&w_{3\,8}^{*}&\sqrt{\frac{2}{3}}w_{0\,8}^{*}&w_{2\,8}^{*}\\
\end{array}
\right), \,\,\,
\widetilde{D}^{\rm opt}_{1\,2}=
\left(
\begin{array}{cccc}
\frac{1}{6}&0&0&0\\
0&w_{5\,7}^{*}&0&w_{4\,7}^{*}\\
0&0&\frac{1}{6}&0\\
0&w_{5\,6}^{*}&0&w_{4\,6}^{*}\\
\end{array}
\right)\,,
\end{equation}
\begin{equation} \label{}
\widetilde{D}^{\rm opt}_{1\,3}=
\left(
\begin{array}{cccc}
0&0&\frac{1}{3}&0\\
0&w_{3\,7}^{*}&0&w_{2\,7}^{*}\\
0&\frac{w_{1\,3}}{\sqrt{6}}&0&\frac{w_{1\,2}}{\sqrt{6}}\\
0&w_{3\,6}^{*}&0&w_{2\,6}^{*}\\
\end{array}
\right), \,\,\,
\widetilde{D}^{\rm opt}_{2\,3}=
\left(
\begin{array}{cccc}
0&0&\frac{1}{3}&0\\
0&w_{3\,5}^{*}&0&w_{2\,5}^{*}\\
0&\frac{w_{1\,3}}{\sqrt{6}}&0&\frac{w_{1\,2}}{\sqrt{6}}\\
0&w_{3\,4}^{*}&0&w_{2\,4}^{*}\\
\end{array}
\right)\,.
\end{equation}
They depend on 64 real parameters, which, when taken small
enough, leave 
$\widetilde{D}^{\rm opt}$ 
positive. We have thus found a large continuous class of symmetric
interference--free 
QCs with maximum average fidelity $5/6$. It turns out that all these
machines are also universal. This is a consequence of imposing zero
interference, as is easily checked by relaxing this condition (see the
discussion of vanishing matrix elements $w_{ij}$ after eq.(\ref{w22})).

\section{Comparison with the Bu\v{z}ek--Hillery QC}
The comparison with the BH QC is not as straightforward as it may seem. The
reason is that BH only specify what happens to a state $|\psi\rangle\otimes
|0\rangle$, but the fate of $|\psi\rangle\otimes
|1\rangle$ remains unspecified. In other words, only part of the dynamical
matrix is defined, and in order to compare our interference--free QCs with the
one by BH  
on the basis of the dynamical matrix,
we would have to extend the definition of the BH QC to deal with the input
state 
$|1\rangle$ for the second qubit as well. If we choose a natural extension,
\begin{eqnarray}
|0\rangle_a|1\rangle_b|Q\rangle_x &\rightarrow& |0\rangle_a |1\rangle_b |
 Q_2\rangle_x + \left( |0\rangle_a |0\rangle_b +|1\rangle_a |1\rangle_b
 \right) |Y_2\rangle_x\nonumber\\
|1\rangle_a|1\rangle_b|Q\rangle_x &\rightarrow& |1\rangle_a |0\rangle_b |
 Q_3\rangle_x + \left( |0\rangle_a |0\rangle_b +|1\rangle_a |1\rangle_b
 \right) |Y_3\rangle_x \,, 
\end{eqnarray}
where the subscript $x$ denotes states of the cloner, 
unitarity imposes the
same constraints as in eq.(3.3) in \cite{Buzek96}, but extended
to $i=0, \ldots ,3$. With an appropriate choice of overlaps between the states
of the cloner, one can show that the BH QC is a member of the continuous class
of QCs 
with dynamical matrices $\widetilde{D}^{\rm opt}$ found above. Alternatively,
we can compare the 
QCs on the basis of the output reduced density matrices $\rho_A'$ and
$\rho_B'$ obtained for mapping states $|\psi\rangle\otimes|0\rangle$. Doing so, 
we find  that {\em all} optimal symmetric interference--free QCs with dynamical
matrices $\widetilde{D}^{\rm opt}$  
give the same single--qubit reduced matrix as the BH QC,   
\begin{equation} \label{}
{\rho_{A}'} ={\rho_{B}'}=
\left(
\begin{array}{cc}
\frac{2}{3}\left|\alpha \right|^2 + \frac{1}{6}&\frac{2}{3}\alpha \beta^{*}\\
\frac{2}{3}\alpha^{*} \beta&\frac{5}{6}- \frac{2}{3}\left|\alpha \right|^2\\
\end{array}
\right)\,,
\end{equation}
with $\rho_{A}'={\rm Tr}_{B}\left[\rho_{AB}' \right]$,
$\rho_{B}'={\rm Tr}_{A}\left[\rho_{AB}' \right] $ and $
\rho_{AB}' = \widetilde{D}^{\rm opt}| \psi \rangle \langle \psi |  \otimes
|0 \rangle\langle 0 |$.

\section{Conclusion}
We have found a large class of 1 $\to$ 2 qubit cloners,
which outperform classical symmetric cloning of qubits in terms of the
average 
fidelities, but do not use any interference in the effective map of the two
qubits. In particular we find QCs which reproduce the optimal average
fidelities 5/6 of symmetric universal QCs without using interference. The
well--known 
Bu\v{z}ek--Hillery QC turns out to be a member of 
this general class of interference--free QCs. How is it possible that
interference--free quantum cloning outperforms classical quantum cloning?
The answer is clear from the structure of the dynamical matrices $D$: Classical
cloning corresponds to a diagonal $D$, quantum cloning to a potentially full
matrix $D$. Interference--free quantum cloning is situated somewhere in
between --- only the off--diagonals of the diagonal blocks need to
vanish. In other words, interference--free propagation of the density matrix
does not map any coherences (i.e. off-diagonal elements of the density
matrix) to probabilities, but may well map 
coherences to other coherences. The latter process influences the
fidelities, and the corresponding matrix elements can be used to optimize
the fidelities beyond classically possible values. It is intriguing
to find that for optimal quantum cloning (in the case of a $1\to 2$ qubit
cloner) just this happens: only probabilities are mapped to probabilities,
whereas coherences never modify the probabilities.
It would be interesting to find out if this generalizes to higher
dimensional or multiple--copy quantum cloning. It is also worthwhile noting
that we have considered here the 
interference in the 
effective propagation of the two qubit system (original and copy). More
general settings could be envisaged, for example including the copying
machine as well. Additional information about the copying
machine is needed then, however, and the results may depend on the dimension
of the Hilbert space of the machine itself.
 
{\em Acknowledgments:} We thank CALMIP (Toulouse) and IDRIS (Orsay) for
the use of their computers. This work was supported by the Agence
National de la Recherche (ANR), project INFOSYSQQ, and the EC
IST-FET project EUROSQIP. B.~Roubert is supported by a grant from the
DGA, with M.~Jacques Blanc--Talon as scientific liaison officer.
\bibliography{../mybibs_bt}

\end{document}